\documentclass[12pt,draft]{article}

\newcommand{\be}{\begin{equation}}
\newcommand{\ee}{\end{equation}}
\newcommand{\bea}{\begin{eqnarray}}
\newcommand{\eea}{\end{eqnarray}}
\newcommand{\beaa}{\begin{eqnarray*}}
\newcommand{\eeaa}{\end{eqnarray*}}
\advance\textheight by \topskip
\usepackage{amssymb}
\usepackage{pstricks}
\newcommand{\M}{\mathcal{M}}
\newcommand{\psl}{\psline[linewidth=0.01]}
\newcommand{\pslv}[1]{\psline[linewidth=#1]}
\newcommand{\one}{\begin{pspicture}(0,.1)(.5,.6)
\psline(0,.25)(.5,.25)
\end{pspicture} }
\newcommand{\ons}{\begin{pspicture}(0,.1)(.5,.6)
\psline(0,.25)(.5,.25)
\pscircle[fillstyle=solid](.25,.25){.08}
\end{pspicture} }
\newcommand{\onsc}{\begin{pspicture}(0,.1)(.5,.6)
\psline(0,.25)(.5,.25)
\pscircle[fillstyle=solid,fillcolor=black](.25,.25){.08}
\end{pspicture} }
\newcommand{\onul}{\begin{pspicture}(0,.1)(.25,.6)
\psline(0,0)(.08,.07)
\psline(.08,.07)(.13,.14)
\psline(.13,.14)(.15,.21)
\psline(.15,.21)(.15,.29)
\psline(.15,.29)(.13,.36)
\psline(.13,.36)(.08,.43)
\psline(.08,.43)(0,.5)
\end{pspicture} }
\newcommand{\onur}{\begin{pspicture}(.25,.1)(.5,.6)
\psline(.5,0)(.42,.07)
\psline(.42,.07)(.37,.14)
\psline(.37,.14)(.35,.21)
\psline(.35,.21)(.35,.29)
\psline(.35,.29)(.37,.36)
\psline(.37,.36)(.42,.43)
\psline(.42,.43)(.5,.5)
\end{pspicture} }
\newcommand{\tws}{\begin{pspicture}(0,.1)(.5,.6)
\psline(0,0)(.5,0)
\psline(0,.5)(.5,.5)
\end{pspicture} }
\newcommand{\twc}{\begin{pspicture}(0,.1)(.5,.6)
\psline(0,0)(.5,.5)
\psline(0,.5)(.5,0)
\end{pspicture} }
\newcommand{\twu}{\onul \onur}
\newcommand{\twssu}{\begin{pspicture}(0,.1)(.5,.6)
\psline(0,0)(.5,0)
\psline(0,.5)(.5,.5)
\pscircle[fillstyle=solid](.25,.5){.08}
\end{pspicture} }
\newcommand{\twssd}{\begin{pspicture}(0,.1)(.5,.6)
\psline(0,0)(.5,0)
\psline(0,.5)(.5,.5)
\pscircle[fillstyle=solid](.25,0){.08}
\end{pspicture} }
\newcommand{\twssb}{\begin{pspicture}(0,.1)(.5,.6)
\psline(0,0)(.5,0)
\psline(0,.5)(.5,.5)
\pscircle[fillstyle=solid](.25,0){.08}
\pscircle[fillstyle=solid](.25,.5){.08}
\end{pspicture} }
\newcommand{\twscsd}{\begin{pspicture}(0,.1)(.5,.6)
\psline(0,0)(.5,0)
\psline(0,.5)(.5,.5)
\pscircle[fillstyle=solid,fillcolor=black](.25,0){.08}
\end{pspicture} }
\newcommand{\res}[2]{\mathop{Res}_{#1}\Big({#2}\Big)}
\newcommand{\ens}{\begin{pspicture}(0,.7)(.5,1.85)
\psline(0,1.75)(.5,1.75)
\psline(0,1.5)(.5,1.5)
\psline(0,1.25)(.5,1.25)
\psline(0,1)(.5,1)
\rput(.25,.85){\vdots}
\psline(0,.5)(.5,.5)
\psline(0,.25)(.5,.25)
\psline(0,0)(.5,0)
\end{pspicture} }
\newcommand{\encu}{\begin{pspicture}(0,.7)(.5,1.85)
\psline(0,1.75)(.5,1.5)
\psline(0,1.5)(.5,1.75)
\psline(0,1.25)(.5,1.25)
\psline(0,1)(.5,1)
\rput(.25,.85){\vdots}
\psline(0,.5)(.5,.5)
\psline(0,.25)(.5,.25)
\psline(0,0)(.5,0)
\end{pspicture} }
\newcommand{\entcu}{\begin{pspicture}(0,.7)(.5,1.85)
\psline(0,1.75)(.5,1.25)
\psline(0,1.5)(.5,1.75)
\psline(0,1.25)(.5,1.5)
\psline(0,1)(.5,1)
\rput(.25,.85){\vdots}
\psline(0,.5)(.5,.5)
\psline(0,.25)(.5,.25)
\psline(0,0)(.5,0)
\end{pspicture} }
\newcommand{\encm}{\begin{pspicture}(0,.7)(.5,1.85)
\psline(0,1.75)(.5,.75)
\psline(0,1.5)(.5,1.75)
\rput(.1,1.25){\vdots}
\rput(.4,1.45){\vdots}
\psline(0,.75)(.5,1)
\psline(0,.5)(.5,.5)
\rput(.25,.35){\vdots}
\psline(0,0)(.5,0)
\end{pspicture} }
\newcommand{\encb}{\begin{pspicture}(0,.7)(.5,1.85)
\psline(0,1.75)(.5,0)
\psline(0,1.5)(.5,1.75)
\psline(0,1.25)(.5,1.5)
\psline(0,1)(.5,1.25)
\rput(.1,.75){\vdots}
\rput(.4,.95){\vdots}
\psline(0,.25)(.5,.5)
\psline(0,0)(.5,.25)
\end{pspicture} }
\newcommand{\enssb}{\begin{pspicture}(0,.7)(.5,1.85)
\psline(0,1.75)(.5,1.75)
\psline(0,1.5)(.5,1.5)
\psline(0,1.25)(.5,1.25)
\psline(0,1)(.5,1)
\rput(.25,.85){\vdots}
\psline(0,.5)(.5,.5)
\psline(0,.25)(.5,.25)
\psline(0,0)(.5,0)
\pscircle[fillstyle=solid](.25,0){.08}
\end{pspicture} }
\newcommand{\ensscu}{\begin{pspicture}(0,.7)(.5,1.85)
\psline(0,1.75)(.5,1.75)
\psline(0,1.5)(.5,1.5)
\psline(0,1.25)(.5,1.25)
\psline(0,1)(.5,1)
\rput(.25,.85){\vdots}
\psline(0,.5)(.5,.5)
\psline(0,.25)(.5,.25)
\psline(0,0)(.5,0)
\pscircle[fillstyle=solid,fillcolor=black](.25,1.75){.08}
\end{pspicture} }
\newcommand{\eni}[1]{\begin{pspicture}(0,.7)(.6,1.85)
\psline(0,1.75)(.6,1.75)
\psline(0,1.5)(.1,1.5)
\psline(.5,1.5)(.6,1.5)
\psline(0,1.25)(.1,1.25)
\psline(.5,1.25)(.6,1.25)
\psline(0,1)(.1,1)
\psline(.5,1)(.6,1)
\psline(.1,0)(.1,1.75)
\rput{90}(.3,.875){{\tiny $#1$}}
\psline(.5,0)(.5,1.75)
\psline(0,.5)(.1,.5)
\psline(.5,.5)(.6,.5)
\psline(0,.25)(.1,.25)
\psline(.5,.25)(.6,.25)
\psline(0,0)(.6,0)
\rput(.05,.85){\vdots}
\rput(.55,.85){\vdots}
\end{pspicture} }
\newcommand{\enin}[1]{\begin{pspicture}(0,.7)(.6,1.85)
\psline(0,1.75)(.6,1.75)
\psline(0,1.5)(.6,1.5)
\psline(0,1.25)(.1,1.25)
\psline(.5,1.25)(.6,1.25)
\psline(0,1)(.1,1)
\psline(.5,1)(.6,1)
\psline(.1,0)(.1,1.5)
\rput{90}(.3,.75){{\tiny $#1$}}
\psline(.5,0)(.5,1.5)
\psline(0,.5)(.1,.5)
\psline(.5,.5)(.6,.5)
\psline(0,.25)(.1,.25)
\psline(.5,.25)(.6,.25)
\psline(0,0)(.6,0)
\rput(.05,.85){\vdots}
\rput(.55,.85){\vdots}
\end{pspicture} }

\newcommand{\entwend}{\begin{pspicture}(-.65,0)(.5,1.4)
\psline(0,.25)(.5,0)
\psline(0,.5)(.5,.25)
\rput(.25,.85){\vdots}
\psline(0,1.25)(.5,1)
\psline(0,1.5)(.5,1.25)
\psline(0,1.75)(.5,1.5)
\psline(0,2)(.5,1.75)
\psline(0,0)(.08,-.035)
\psline(.08,-.035)(.13,-.07)
\psline(.13,-.07)(.15,-.105)
\psline(.15,-.105)(.15,-.145)
\psline(.15,-.145)(.13,-.18)
\psline(.13,-.18)(.08,-.215)
\psline(.08,-.215)(0,-.25)
\end{pspicture} }
\newcommand{\entwist}[4]{\left(\begin{pspicture}(0,.7)(#2,2.1)
\psline(0,0)(.5,-.25)
\psline(0,.25)(.5,0)
\psline(0,.5)(.5,.25)
\rput(.25,.85){\vdots}
\psline(0,1.25)(.5,1)
\psline(0,1.5)(.5,1.25)
\psline(0,1.75)(.5,1.5)
\psline(.5,1.75)(.42,1.785)
\psline(.42,1.785)(.37,1.82)
\psline(.37,1.82)(.35,1.855)
\psline(.35,1.855)(.35,1.895)
\psline(.35,1.895)(.37,1.93)
\psline(.37,1.93)(.42,1.965)
\psline(.42,1.965)(.5,2)
\rput(#1,.7){\entwend}
\psline(.5,-.25)(#1,-.25)
\psline(.5,2)(#1,2)
\rput(#3,.825){#4}
\end{pspicture}\right)}
\makeatletter

\@addtoreset{equation}{section}

\def\section{\@startsection{section}{1}{\z@}{-3.5ex plus -1ex minus
 -.2ex}{2.3ex plus .2ex}{\large\bf\centering}}
\def\subsection{\@startsection{subsection}{2}{\z@}{-3.25ex plus%
 -1ex minus -.2ex}{1.5ex plus .2ex}{\bf}}
\def\subsubsection{\@startsection{subsubsection}{3}{\z@}{-3.25ex plus%
 -1ex minus -.2ex}{1.5ex plus .2ex}{\sl}}
\makeatother
\begin{document}
\baselineskip 18pt
\parindent 12pt
\begin{flushright}
hep-th/0204072\\
April 2002\\[3mm]
\end{flushright}
\vspace{.4cm}
\begin{center}
{\Large {\bf Boundary scattering in the $SU(N)$ \\
principal chiral model on the half-line \\
with conjugating boundary conditions}}\\ \vspace{1cm} {\large
B. J. Short\footnote{\tt bjs108@york.ac.uk}}
\\
\vspace{3mm} {\em Department of Mathematics,\\ University of York,
\\York YO10 5DD, U.K.}
\end{center}
\vspace{1ex}
\begin{abstract}
\noindent We investigate the $SU(N)$ Principal Chiral Model on a
half-line with a particular set of boundary conditions (BCs). In
previous work these BCs have been shown to correspond to boundary
scattering matrices ($K$-matrices) which are representation conjugating
and whose matrix structure corresponds to one of the symmetric spaces
$SU(N)/SO(N)$ or $SU(N)/S\!p(N)$. Starting from the bulk particle
spectrum and the $K$-matrix for a particle in the vector representation
we construct $K$-matrices for particles in higher rank representations
scattering off the boundary. We then perform an analysis of the physical
strip pole structure and provide a complete set of boundary Coleman-Thun
mechanisms for those poles which do not correspond to particles coupling
to the boundary. We find that the model has no non-trivial boundary
states.
\end{abstract}
\vspace{1ex}

\section{The bulk PCM and boundary scattering in the vector
representation}
\label{sec:bbv}

The principal chiral field, $g(x,t)$, takes values in a compact Lie
group $G$. Its dynamics are governed by the lagrangian
\be
\mathcal{L}=\frac{1}{2}{\rm Tr}\Big(\partial_{\mu}g^{-1}\partial^{\mu}g
\Big).
\ee
In the bulk model the spacetime coordinates $x,t$ are allowed to take
any values in the reals, $-\infty<x,t<\infty$.

The model is classically integrable~\cite{lusch78,berna91,evans99} and
this classical integrability is believed to extend to the quantum level,
allowing the techniques of 1+1 dimensional integrable QFT to be applied.
The exact $S$-matrices describing two-particle interactions were
calculated for various cases by Ogievetsky, Reshetikhin and
Wiegmann~\cite{ogiev87}.

In this paper we shall be concerned with the particular case $G=SU(N)$,
for which the $S$-matrix describing the scattering of two vector
particles can be written
\be
S^{PCM}_{(1,1)}(\theta)=X_{(1,1)}(\theta)\Big({S_{(1,1)}(\theta)}_L
\otimes{S_{(1,1)}(\theta)}_R\Big).
\ee
Here $X_{(1,1)}(\theta)=(4)$ is the CDD factor, where
\be
(y)=\frac{\sinh{(\frac{\theta}{2}+\frac{i\pi y}{4N})}}{\sinh{(\frac{
\theta}{2}-\frac{i\pi y}{4N})}}
\ee
and ${S_{(1,1)}(\theta)}_{L,R}$ are left and right copies of the minimal
$S$-matrix
\be
\label{eqn:msm}
S_{(1,1)}(\theta)=\sigma(\theta)\mathbb{P}\Big(P_2^S-[4]P_2^A\Big).
\ee
The scalar prefactor $\sigma(\theta)$ is given by
\be
\sigma(\theta)=\frac{\Gamma\left(\frac{\theta}{2i\pi}+\frac{1}{N}\right)
\Gamma\left(\frac{-\theta}{2i\pi}\right)}{\Gamma\left(\frac{-\theta}
{2i\pi}+\frac{1}{N}\right)\Gamma\left(\frac{\theta}{2i\pi}\right)},
\ee
whilst the square brackets denote the function
\be
[y]=\frac{2N\theta+i\pi y}{2N\theta-i\pi y}.
\ee
The $S$-matrix~(\ref{eqn:msm}) acts on the tensor product of two copies
of the vector representation space: $\mathbb{P}$ is the transposition
operator acting on the two spaces, whilst $P_2^{S,A}$ are the projectors
onto the symmetric and antisymmetric subspaces of the product space. The
variable $\theta$ is the rapidity difference between the two incoming
multiplets; for a discussion of 1+1 dimensional factorisable scattering
theory the reader is directed to~\cite{dorey98}.

\subsection{The spectrum of bound states}
\label{subsec:sbs}

The bound state spectrum of the bulk model was described
in~\cite{ogiev87}; here we shall recall some of the features of the
particles and their interactions which will prove useful later.

Firstly, there are $N\!-\!1$ particle types, which transform as the
fundamental representations (that is the completely antisymmetric
representations) of $SU(N)$. We label the particle type transforming as
the $n$th fundamental representation by $n$. Thus a particle of type $n$
is conjugate to one of type $N\!-\!n$.

If we set the mass scale $\M$ to be such that the mass of the vector
particle is given by
$$\M_1=\M\sin\left(\frac{\pi}{N}\right)$$
then the mass of the $n$th particle is given by
\be
\M_n=\M\sin\left(\frac{n\pi}{N}\right).
\ee
This mass formula follows from the fact that the particle interactions
are such that particles of type $n$ and $m$ (for $n\!+\!m\leq N$)
scattering at rapidity difference $\theta=\frac{(n+m)i\pi}{N}$ form a
particle of type $n\!+\!m$ as a bound state.

\subsection{The half-line model: scattering of the vector particle}
\label{subsec:hsv}

We now turn our attention to the PCM on the half-line: the space
coordinate is restricted to the range $-\infty<x\leq 0$. In a previous
paper~\cite{macka01} various boundary conditions were found which are
believed to preserve the quantum integrability of the model with
boundary.

The $S$-matrix description for the model must now be extended to include
scattering off the boundary. It was established in~\cite{macka01} that
certain classical boundary conditions classified by the symmetric spaces
$$\frac{SU(N)}{SO(N)}\qquad{\rm and}\qquad\frac{SU(N)}{S\!p(N)}$$
correspond to representation conjugating $K$-matrices. These
$K$-matrices, describing the scattering of a vector particle off the
boundary, are given by
\be
K^{PCM}_1(\theta)=Y_1(\theta)\Big({K_1(\theta)}_L\otimes{K_1(\theta)}_R\Big). 
\ee
The CDD factor is given by
\be
Y_1(\theta)=(\gamma N\!+\!2)(\delta N\!+\!4),
\ee
where the four choices $\gamma,\delta=1,3$ each provide a suitable PCM
$K$-matrix for the vector particle: in each of the four cases the
$K$-matrix has no poles in the physical strip
(Im$(\theta)\in[0,\frac{\pi}{2}]$). We shall find in
section~\ref{sec:pkb} that one of the four choices is preferred by the
boundary interactions of the higher rank particles.
${K_1(\theta)}_{L,R}$ are left and right copies of the minimal
$K$-matrices
\be
K_1(\theta)=\rho(\theta)E
\ee
where the matrix part, $E$, is unitary; it is symmetric in the case of
$SU(N)/SO(N)$ and antisymmetric in the case of $SU(N)/S\!p(N)$. The
scalar prefactor is given by
\be
\rho(\theta)=\frac{\Gamma\left(\frac{\theta}{2i\pi}\!+\!\frac{1}
{4}\right)\Gamma\left(\frac{-\theta}{2i\pi}\!+\!\frac{\alpha}{4}\!+\!
\frac{1}{2N}\right)}{\Gamma\left(\frac{-\theta}{2i\pi}\!+\!\frac{1}
{4}\right)\Gamma\left(\frac{\theta}{2i\pi}\!+\!\frac{\alpha}{4}\!+\!
\frac{1}{2N}\right)}
\ee
where $\alpha=1$ for $SU(N)/SO(N)$ and $\alpha=3$ for $SU(N)/S\!p(N)$.

Before going on to consider the scattering of higher rank bulk particles
off the ground state, we note that because neither of these PCM
$K$-matrices has any poles on the physical strip the bulk vector
particle cannot bind to the boundary with either choice of BC.

\section{Higher rank particle boundary scattering}

We calculate the $K$-matrices for higher rank particles by fusion. We
shall illustrate the procedure, in some detail, with the $n=2$ particle
before proceeding to the general case. Diagrammatically this calculation
is \\
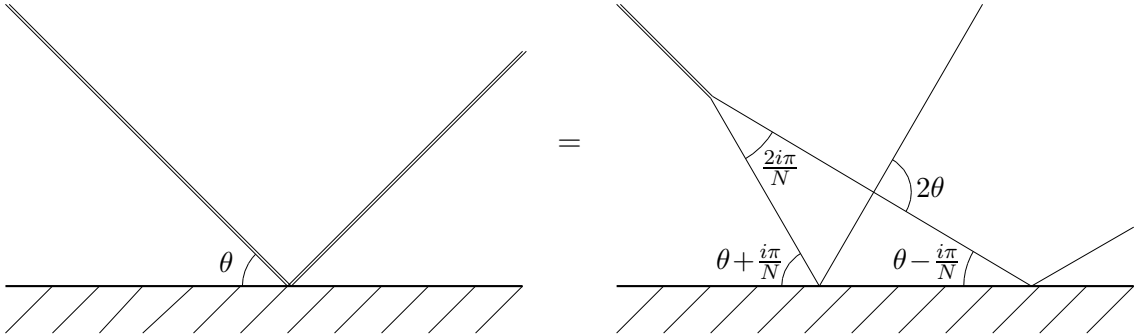
\begin{pspicture}(-.55,0)(15,4.875)
\psline(0,.625)(6.875,.625)
\psline(8.125,.625)(15,.625)
\psl(0,0)(.625,.625)
\psl(.625,0)(1.25,.625)
\psl(1.25,0)(1.825,.625)
\psl(1.825,0)(2.5,.625)
\psl(2.5,0)(3.125,.625)
\psl(3.125,0)(3.75,.625)
\psl(3.75,0)(4.375,.625)
\psl(4.375,0)(5,.625)
\psl(5,0)(5.625,.625)
\psl(5.625,0)(6.25,.625)
\psl(6.25,0)(6.875,.625)
\psl(8.125,0)(8.75,.625)
\psl(8.75,0)(9.375,.625)
\psl(9.375,0)(10,.625)
\psl(10,0)(10.625,.625)
\psl(10.625,0)(11.25,.625)
\psl(11.25,0)(11.875,.625)
\psl(11.875,0)(12.5,.625)
\psl(12.5,0)(13.125,.625)
\psl(13.125,0)(13.75,.625)
\psl(13.75,0)(14.375,.625)
\psl(14.375,0)(15,.625)
\psl(0,4.375)(3.75,.625)
\psl(.05,4.375)(3.8,.625)
\psl(3.75,.625)(6.875,3.75)
\psl(3.8,.625)(6.925,3.75)
\psl(3.15,.625)(3.1584,.725)
\psl(3.1584,.725)(3.1843,.825)
\psl(3.1843,.825)(3.2304,.925)
\psl(3.2304,.925)(3.2816,1)
\psl(3.2816,1)(3.3257,1.0493)
\rput(2.938,.95){{\small $\theta$}}
\psl(8.125,4.375)(9.375,3.125)
\psl(8.175,4.375)(9.4,3.15)
\psl(9.375,3.125)(10.82,.625)
\psl(9.4,3.15)(13.64,.625)
\psl(10.82,.625)(12.99,4.375)
\psl(13.64,.625)(15,1.41)
\psl(10.0453,2.4797)(9.9382,2.3868)
\psl(10.0453,2.4797)(10.1382,2.5868)
\psl(9.9382,2.3868)(9.8386,2.321)
\psl(10.1382,2.5868)(10.1974,2.6751)
\rput(10.3,2.2){{\small $\frac{2i\pi}{N}$}}
\psl(10.32,.625)(10.3301,.725)
\psl(10.3301,.725)(10.3617,.825)
\psl(10.3617,.825)(10.42,.925)
\psl(10.42,.925)(10.52,1.025)
\psl(10.52,1.025)(10.5698,1.0579)
\rput(9.9,.95){{\small $\theta\!+\!\frac{i\pi}{N}$}}
\psl(12.74,.625)(12.7456,.725)
\psl(12.7456,.725)(12.7625,.825)
\psl(12.7625,.825)(12.7915,.925)
\psl(12.7915,.925)(12.8338,1.025)
\psl(12.8338,1.025)(12.8667,1.0855)
\rput(12.25,.95){{\small $\theta\!-\!\frac{i\pi}{N}$}}
\psl(12.0427,1.874)(12.0326,1.974)
\psl(12.0427,1.874)(12.0326,1.774)
\psl(12.0326,1.974)(12.001,2.074)
\psl(12.0326,1.774)(12.001,1.674)
\psl(12.001,2.074)(11.9427,2.174)
\psl(12.001,1.674)(11.9723,1.6182)
\psl(11.9427,2.174)(11.8427,2.274)
\psl(11.8427,2.274)(11.7931,2.3068)
\rput(12.3,1.95){{\small $2\theta$}}
\rput(7.5,2.5){$=$}
\end{pspicture} \\
Thus, we have the following expression for the second rank particle
minimal $K$-matrix
\be
K_2(\theta)=S_{(1,1)}({\textstyle\frac{2i\pi}{N}})\left(I\otimes K_1(
\theta\!+\!{\textstyle\frac{i\pi}{N}})\right)S_{(1,\bar{1})}(2\theta)
\left(I\otimes K_1(\theta\!-\!{\textstyle\frac{i\pi}{N}})\right).
\ee
We note that since the reflection matrix is representation conjugating
it is $S_{(1,\bar{1})}(2\theta)$ that we require. This is obtained as
\be
S_{(1,\bar{1})}(\theta)=\omega(i\pi\!-\!\theta)\left(\twu-{\textstyle
\frac{N(i\pi-\theta)}{2i\pi}}\twc\right),\qquad\left({\rm Note\colon}\ 
\omega(\theta)=\frac{\sigma(\theta)}{(1\!-\!\frac{N\theta}{2i\pi})}
\right)
\ee
where we have introduced a diagrammatic notation to make the subsequent
calculations more transparent: we represent the $N\times N$ identity
matrix by a line, $\one$. We denote the matrix part, $E$, of the vector
particle $K$-matrix by $\ons$. Matrix multiplication is given by
concatenation of the lines and the scattering order goes from right to
left. The lines are implicitly directed, with conjugation of the
particle reversing the line direction, but we can ignore this here.

Substituting in for all the minimal $S$- and $K$-matrices we obtain
\be
K_2(\theta)=\omega({\textstyle\frac{2i\pi}{N}})\omega(i\pi\!-\!2\theta
)\rho(\theta\!+\!{\textstyle\frac{i\pi}{N}})\rho(\theta\!-\!{\textstyle
\frac{i\pi}{N}})\left(\tws-\twc\right)\twssd\left(\twu-{\textstyle\frac{
N(i\pi-2\theta)}{2i\pi}}\twc\right)\twssd\,.
\ee
At this point the cases $SU(N)/SO(N)$ and $SU(N)/S\!p(N)$ diverge: we
need to consider the matrix parts separately. Firstly we consider
$SU(N)/SO(N)$, for which $(\ons)^t=\ons$. We have
\begin{eqnarray*}
\left(\tws-\twc\right)\twssd\left(\twu-{\textstyle\frac{N(i\pi-2\theta)}
{2i\pi}}\twc\right)\twssd&=&{\textstyle\frac{N(i\pi-2\theta)}{2i\pi}}
\left(\tws-\twc\right)\twssb \\
&=&{\textstyle\frac{N(i\pi-2\theta)}{i\pi}}P_2\,,
\end{eqnarray*}
where we have set
\be
P_2={\textstyle\frac{1}{2}}\left(\tws-\twc\right)\twssb\,.
\ee
This satisfies $P_2(P_2)^{\dag}=P_2^A$ (we recall that $E$ is unitary).

Moving on to the case $SU(N)/S\!p(N)$, for which $(\ons)^t=-(\ons)$, we
have
\begin{eqnarray*}
\left(\tws-\twc\right)\twssd\left(\twu-{\textstyle\frac{N(i\pi-2\theta)}
{2i\pi}}\twc\right)\twssd&=&2\twssu\twu\twssu+{\textstyle\frac{N(i\pi-2
\theta)}{2i\pi}}\left(\tws-\twc\right)\twssb \\
&=&{\textstyle\frac{N(i\pi-2\theta)}{i\pi}}\Big(P_2^{(2)}-\left[N\right]
P_2^{(0)}\Big),
\end{eqnarray*}
where we have set
\be
P_2^{(2)}={\textstyle\frac{1}{2}}\left(\tws-\twc\right)\twssb+
{\textstyle\frac{1}{N}}\twssd\twu\twssd\,,\qquad P_2^{(0)}={\textstyle
\frac{1}{N}}\twssd\twu\twssd\,.
\ee
Setting $(\ons)^{\dag}=\onsc$, these satisfy (recall $\ons\onsc=\one$)
\be
P_2^{(2)}(P_2^{(2)})^{\dag}={\textstyle\frac{1}{2}}\left(\tws-\twc
\right)-{\textstyle\frac{1}{N}}\twssd\twu\twscsd\,,\qquad P_2^{(0)}(
P_2^{(0)})^{\dag}={\textstyle\frac{1}{N}}\twssd\twu\twscsd\,,
\ee
which are orthogonal projectors.

In both cases, $SU(N)/SO(N)$ and $SU(N)/S\!p(N)$, we obtain the scalar
prefactor
\be
\rho_2(\theta)=(N\!-\!{\textstyle\frac{2N\theta}{i\pi}})\omega(
{\textstyle\frac{2i\pi}{N}})\omega(i\pi\!-\!2\theta)\rho(\theta\!+\!
{\textstyle\frac{i\pi}{N}})\rho(\theta\!-\!{\textstyle\frac{i\pi}{N}}).
\ee
Thus we have calculated the second rank particle minimal $K$-matrices
\bea
K_2(\theta)=\rho_2(\theta)P_2&&SU(N)/SO(N), \\
K_2(\theta)=\rho_2(\theta)\Big(P_2^{(2)}-\left[N\right]P_2^{(0)}\Big)
&&SU(N)/S\!p(N).
\eea
The CDD factor is given by
\be
Y_2(\theta)=\res{v=\frac{2i\pi}{N}}{X_{(1,1)}(v)}Y_1(\theta\!+\!
{\textstyle\frac{i\pi}{N}})X_{(1,1)}(i\pi\!-\!2\theta)Y_1(\theta\!-\!
{\textstyle\frac{i\pi}{N}}),
\ee
and so we have the PCM $K$-matrices
\be
K^{PCM}_2(\theta)=Y_2(\theta)\Big({K_2(\theta)}_L\otimes{K_2(\theta)}_R\Big). 
\ee

\subsection{The $n$th rank particle $K$-matrices}

Having considered the calculation of the second rank particle
$K$-matrices, we now proceed to the general $n$th rank case. We consider
the diagram \\
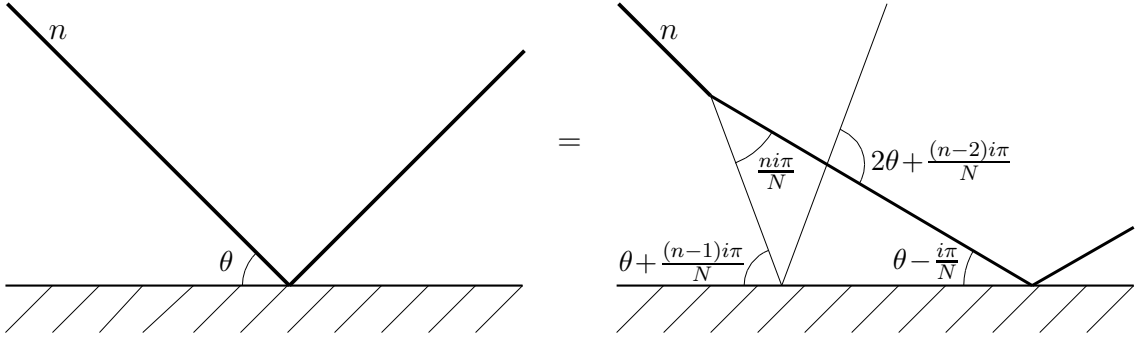
\begin{pspicture}(-.55,0)(15,4.875)
\psline(0,.625)(6.875,.625)
\psline(8.125,.625)(15,.625)
\psl(0,0)(.625,.625)
\psl(.625,0)(1.25,.625)
\psl(1.25,0)(1.825,.625)
\psl(1.825,0)(2.5,.625)
\psl(2.5,0)(3.125,.625)
\psl(3.125,0)(3.75,.625)
\psl(3.75,0)(4.375,.625)
\psl(4.375,0)(5,.625)
\psl(5,0)(5.625,.625)
\psl(5.625,0)(6.25,.625)
\psl(6.25,0)(6.875,.625)
\psl(8.125,0)(8.75,.625)
\psl(8.75,0)(9.375,.625)
\psl(9.375,0)(10,.625)
\psl(10,0)(10.625,.625)
\psl(10.625,0)(11.25,.625)
\psl(11.25,0)(11.875,.625)
\psl(11.875,0)(12.5,.625)
\psl(12.5,0)(13.125,.625)
\psl(13.125,0)(13.75,.625)
\psl(13.75,0)(14.375,.625)
\psl(14.375,0)(15,.625)
\pslv{.05}(.025,4.375)(3.775,.625)
\pslv{.05}(3.775,.625)(6.9,3.75)
\psl(3.15,.625)(3.1584,.725)
\psl(3.1584,.725)(3.1843,.825)
\psl(3.1843,.825)(3.2304,.925)
\psl(3.2304,.925)(3.2816,1)
\psl(3.2816,1)(3.3257,1.0493)
\rput(2.938,.95){{\small $\theta$}}
\pslv{.05}(8.15,4.375)(9.375,3.15)
\psl(9.375,3.15)(10.32,.625)
\pslv{.04}(9.375,3.15)(13.65,.625)
\psl(10.32,.625)(11.72,4.375)
\pslv{.04}(13.65,.625)(15,1.4)
\psl(10.0453,2.4797)(9.9382,2.3868)
\psl(10.0453,2.4797)(10.1382,2.5868)
\psl(9.9382,2.3868)(9.8386,2.321)
\psl(10.1382,2.5868)(10.1974,2.6751)
\psl(9.8386,2.321)(9.7,2.255)
\rput(10.25,2.15){{\small $\frac{ni\pi}{N}$}}
\psl(9.82,.625)(9.8301,.725)
\psl(9.8301,.725)(9.8617,.825)
\psl(9.8617,.825)(9.92,.925)
\psl(9.92,.925)(10.02,1.025)
\psl(10.02,1.025)(10.08,1.0636)
\psl(10.08,1.0636)(10.1447,1.0933)
\rput(9,.95){{\small $\theta\!+\!\frac{(n-1)i\pi}{N}$}}
\psl(12.74,.625)(12.7456,.725)
\psl(12.7456,.725)(12.7625,.825)
\psl(12.7625,.825)(12.7915,.925)
\psl(12.7915,.925)(12.8338,1.025)
\psl(12.8338,1.025)(12.8667,1.0855)
\rput(12.25,.95){{\small $\theta\!-\!\frac{i\pi}{N}$}}
\psl(11.4217,2.2369)(11.4116,2.1369)
\psl(11.4217,2.2369)(11.4116,2.3369)
\psl(11.4116,2.1369)(11.38,2.0369)
\psl(11.4116,2.3369)(11.38,2.4369)
\psl(11.38,2.0369)(11.3523,1.9826)
\psl(11.38,2.4369)(11.3217,2.5369)
\psl(11.3217,2.5369)(11.2127,2.6369)
\psl(11.2127,2.6369)(11.0966,2.7053)
\rput(12.45,2.3){{\small $2\theta\!+\!\frac{(n-2)i\pi}{N}$}}
\rput(7.5,2.5){$=$}
\multirput(.7,4)(8.125,0){2}{$n$}
\end{pspicture} \\
Algebraically we have
\bea
K_n(\theta)=S_{(1,n-1)}({\textstyle\frac{ni\pi}{N}})\Big(I^{\otimes(n-1)
}\otimes K_1(\theta\!+\!{\textstyle\frac{(n-1)i\pi}{N}})\Big) \nonumber
\\
\tilde{S}_{(n-1,\bar{1})}(2\theta\!+\!{\textstyle\frac{(n-2)i\pi}{N}})
\Big(I\otimes\tilde{K}_{n-1}(\theta\!-\!{\textstyle\frac{i\pi}{N}})\Big),
\eea
where the relations
\bea
S_{(n-1,\bar{1})}(\phi)=\Omega(n\!-\!1)\tilde{S}_{(n-1,\bar{1})}(\phi),
\qquad K_{n-1}(\phi)=\Omega(n\!-\!1)\tilde{K}_{n-1}(\phi) \\
{\rm and}\quad\Omega(n)=\prod_{l=1}^{n-1}{\prod_{k=1}^{l}{(k\!+\!1)
\omega({\textstyle\frac{2ki\pi}{N}})}}
\eea
define the quantities $\tilde{S}$ and $\tilde{K}$. (Note that we keep
track of these $\theta$-independent factors, $\Omega$, for the sake of
completeness rather than necessity.) 

In order to calculate $K_n(\theta)$ from the expression above we again
turn to our diagrammatic notation. We can then express the $S$-matrix
$\tilde{S}_{(1,n-1)}(\theta)$ as
\be
\tilde{S}_{(1,n-1)}(\theta)=\psi_{n-1}(\theta)\enin{P_{n\!-\!1
}^A}\left(\ens-\encu+\entcu+\dots+(-1)^{n-1}({\textstyle\frac{N\theta}
{2i\pi}}\!-\!{\textstyle\frac{n}{2}}\!+\!1)\encb\right),
\ee
where
\be
\psi_{n-1}(\theta)=\left(\prod_{k=1}^{n-1}{\omega(\theta\!-\!
{\textstyle\frac{ni\pi}{N}}\!+\!{\textstyle\frac{2ki\pi}{N}})}\right)
\left(\prod_{k=2}^{n-1}{({\textstyle\frac{N\theta}{2i\pi}}\!-\!
{\textstyle\frac{n}{2}}\!+\!k)}\right)
\ee
and $P_{n-1}^A$ is the projector onto the completely antisymmetric
subspace of the $n\!-\!1$-fold tensor product of vector representations
(the $n\!-\!1$\,th fundamental representation of $SU(N)$). Setting
$\theta=\frac{ni\pi}{N}$ we obtain
\be
\tilde{S}_{(1,n-1)}({\textstyle\frac{ni\pi}{N}})=n\psi_{n-1}({\textstyle
\frac{ni\pi}{N}})P_n^A\quad\Longrightarrow\quad S_{(1,n-1)}({\textstyle
\frac{ni\pi}{N}})=\Omega(n)P_n^A,
\ee
which follows from $n\Omega(n\!-\!1)\psi_{n-1}(\frac{ni\pi}{N})=\Omega(n
)$. We note that the above expression for $\tilde{S}_{(1,n-1)}(\theta)$
is obtained by fusing bulk vector particles together. We will not prove
the result here (it was established in~\cite{ogiev87}), but remark that
we have provided enough details for the interested reader to assemble an
inductive proof fairly straightforwardly.

In order to obtain $\tilde{S}_{(n-1,\bar{1})}(\theta)$ we perform the
usual crossing operation, obtaining
\bea
\tilde{S}_{(n-1,\bar{1})}(\theta)=\psi_{n-1}(i\pi\!-\!\theta)
\nonumber \\
\entwist{11.1}{11.63}{5.8}{$\left(\enin{P_{n\!-\!1}^A}\right)\!\!\!
\left(\ens-\encu+\entcu+\dots+(-1)^{n-1}({\textstyle\frac{N}{2}}\!-\!
{\textstyle\frac{N\theta}{2i\pi}}\!-\!{\textstyle\frac{n}{2}}\!+\!1)
\encb\right)$}.
\eea

Substituting into our expression for $K_n(\theta)$ we have
\bea
\label{knt}
K_n(\theta)=\Omega(n)\rho(\theta\!+\!{\textstyle\frac{(n-1)i\pi}{N}})
\tilde{\rho}_{n-1}(\theta\!-\!{\textstyle\frac{i\pi}{N}})\psi_{n-1}(i\pi
\!-\!2\theta\!+\!{\textstyle\frac{(2-n)i\pi}{N}}) \nonumber \\
\eni{P_n^A}\enssb\entwist{11.1}{11.63}{5.8}{$\left(\enin{P_{n\!-\!1}^A}
\right)\!\!\!\left(\ens-\encu+\entcu+\dots+(-1)^{n-1}({\textstyle\frac{N
}{2}}\!-\!{\textstyle\frac{N\theta}{i\pi}}\!-\!n\!+\!2)\encb\right)$}
\enin{X_{n\!-\!1}(\theta\!\!-\!\!{\textstyle\frac{i\pi}{N}})}\,,
\eea
where we have split the $n\!-\!1$ rank particle $K$-matrix into a scalar
prefactor and a matrix part,
\be
\tilde{K}_{n-1}(\theta)=\tilde{\rho}_{n-1}(\theta)X_{n-1}(\theta).
\ee
We again have to consider the two cases separately, as we did for $n=2$;
firstly we calculate the minimal $K$-matrix for $SU(N)/SO(N)$.

\subsection{Calculating $K_n(\theta)$ for $SU(N)/SO(N)$}

In the case of $SU(N)/SO(N)$ we shall find that the matrix part of the
minimal $K$-matrix, $X_n(\theta)$, is constant. We recall that
$X_1(\theta)=E(=P_1)$ and $X_2(\theta)=P_2$ and we take as an induction
hypothesis $X_{n-1}(\theta)=P_{n-1}$ (with $P_{n-1}$ the obvious
generalisation of $P_1$ and $P_2$; thus $P_{n-1}(P_{n-1})^{\dag}=P_{n-1
}^A$). Substituting into our expression for $K_n(\theta)$ we have
\bea
K_n(\theta)=\Omega(n)\rho(\theta\!+\!{\textstyle\frac{(n-1)i\pi}{N}})
\tilde{\rho}_{n-1}(\theta\!-\!{\textstyle\frac{i\pi}{N}})\psi_{n-1}(
i\pi\!-\!2\theta\!+\!{\textstyle\frac{(2-n)i\pi}{N}}) \nonumber \\
\eni{P_n^A}\enssb\entwist{11.1}{11.63}{5.8}{$\left(\enin{P_{n\!-\!1}^A}
\right)\!\!\!\left(\ens-\encu+\entcu+\dots+(-1)^{n-1}({\textstyle\frac{N
}{2}}\!-\!{\textstyle\frac{N\theta}{i\pi}}\!-\!n\!+\!2)\encb\right)$}
\enin{P_{n\!-\!1}}\,.
\eea
The $P_{n-1}^A$ factor contracts with the $P_n^A$ factor, leaving us to
consider a sum of terms whose matrix parts are of the form
$$\eni{P_n^A}\enssb\entwist{1.7}{2.23}{1.1}{$\left(\encm\right)$}\enin{
P_{n\!-\!1}}\,.$$
Since $\ons$ is symmetric it is clear that all these will give zero
except the contribution from
$$\eni{P_n^A}\enssb\entwist{1.7}{2.23}{1.1}{$\left(\encb\right)$}\enin{
P_{n\!-\!1}}=(-1)^{n-1}\eni{P_n}\,.$$
Thus, we have
\be
K_n(\theta)=\Omega(n)\rho(\theta\!+\!{\textstyle\frac{(n-1)i\pi}{N}})
\tilde{\rho}_{n-1}(\theta\!-\!{\textstyle\frac{i\pi}{N}})\psi_{n-1}
(i\pi\!-\!2\theta\!+\!{\textstyle\frac{(2-n)i\pi}{N}})({\textstyle\frac{
N}{2}}\!-\!{\textstyle\frac{N\theta}{i\pi}}\!-\!n\!+\!2)P_n.
\ee
From this we read off the scalar prefactor
\be
\rho_n(\theta)=\Omega(n)\rho(\theta\!+\!{\textstyle\frac{(n-1)i\pi}{N}})
\tilde{\rho}_{n-1}(\theta\!-\!{\textstyle\frac{i\pi}{N}})\psi_{n-1}(i\pi
\!-\!2\theta\!+\!{\textstyle\frac{(2-n)i\pi}{N}})({\textstyle\frac{N}{2}
}\!-\!{\textstyle\frac{N\theta}{i\pi}}\!-\!n\!+\!2),
\ee
whilst the matrix part is $X_n(\theta)=P_n$. We have proven the
conjectured form of $K_n(\theta)$ by induction and established a
recurrence relation for the scalar prefactor, $\rho_n(\theta)$. We turn
our attention now to the analogous calculation for $SU(N)/S\!p(N)$.

\subsection{Calculating $K_n(\theta)$ for $SU(N)/S\!p(N)$}

For the $SU(N)/S\!p(N)$ case we appeal to the boundary tensor product
graph method~\cite{deliu01}, which gives the matrix part of the minimal
$K$-matrix as
\be
X_n(\theta)=\sum_{r=0}^{\lfloor\frac{n}{2}\rfloor}{(-1)^r\prod_{k=1}^{r}
{[N\!-\!2n\!+\!4k]}P_n^{(n-2r)}},
\ee
where $P_n^{(n-2r)}(P_n^{(n-2r)})^{\dag}$ are the orthogonal projectors
corresponding to the irreducible representations of the embedded
$S\!p(N)$. We note the relation $P_n^{(n-2r)}(P_n^{(n-2s)})^{\dag}=0$
for $r\neq s$, which allows us to write
\be
\label{kntr}
K_n(\theta)(P_n^{(n)})^{\dag}=\rho_n(\theta)P_n^{(n)}(P_n^{(n)})^{\dag}.
\ee

If we are to simplify our expression for $K_n(\theta)$~(\ref{knt}) we
will need to know how the operators $P_{n-1}^{(n-1-2r)}$, in $K_{n-1}(
\theta\!-\!\frac{i\pi}{N})$, relate to the operators $P_n^{(n-2s)}$. The
relation which allows us to make progress is $(E\otimes P_{n-1}^{(n-1)})
(P_n^{(n)})^{\dag}=P_n^{(n)}(P_n^{(n)})^{\dag}$; then we have
\bea
K_n(\theta)(P_n^{(n)})^{\dag}=\Omega(n)\rho(\theta\!+\!{\textstyle\frac{
(n-1)i\pi}{N}})\tilde{\rho}_{n-1}(\theta\!-\!{\textstyle\frac{i\pi}{N}})
\psi_{n-1}(i\pi\!-\!2\theta\!+\!{\textstyle\frac{(2-n)i\pi}{N}})
\nonumber \\
\eni{P_n^A}\enssb\entwist{11.1}{11.63}{5.8}{$\left(\enin{P_{n\!-\!1}^A}
\right)\!\!\!\left(\ens-\encu+\entcu+\dots+(-1)^{n-1}({\textstyle\frac{N
}{2}}\!-\!{\textstyle\frac{N\theta}{i\pi}}\!-\!n\!+\!2)\encb\right)$}
\ensscu\eni{P_n^{(n)}(P_n^{(n)})^{\dag}}\,.
\eea
The $P_{n-1}^A$ factor contracts with the $P_n^A$ factor, leaving us to
consider a sum of terms whose matrix parts are of the form
$$\eni{P_n^A}\enssb\entwist{1.7}{2.23}{1.1}{$\left(\encm\right)$}\ensscu
\eni{P_n^{(n)}(P_n^{(n)})^{\dag}}\,.$$
The orthogonal property, $(P_n^{(n-2r)})^{\dag}P_n^{(n)}=0$ for $r\neq
0$, guarantees that the only such term not equal to zero is
$$\eni{P_n^A}\enssb\entwist{1.7}{2.23}{1.1}{$\left(\encb\right)$}\ensscu
\eni{P_n^{(n)}(P_n^{(n)})^{\dag}}=(-1)^{n-1}\eni{P_n^{(n)}(P_n^{(n)})^{
\dag}}\,.$$
Thus, we have
\be
K_n(\theta)(P_n^{(n)})^{\dag}\!=\Omega(n)\rho(\theta\!+\!{\textstyle
\frac{(n-1)i\pi}{N}})\tilde{\rho}_{n-1}(\theta\!-\!{\textstyle\frac{i\pi
}{N}})\psi_{n-1}(i\pi\!-\!2\theta\!+\!{\textstyle\frac{(2-n)i\pi}{N}})({
\textstyle\frac{N}{2}}\!-\!{\textstyle\frac{N\theta}{i\pi}}\!-\!n\!+\!2)
P_n^{(n)}(P_n^{(n)})^{\dag}.
\ee
Comparing this with the expression~(\ref{kntr}) we obtain
\be
\rho_n(\theta)=\Omega(n)\rho(\theta\!+\!{\textstyle\frac{(n-1)i\pi}{N}})
\tilde{\rho}_{n-1}(\theta\!-\!{\textstyle\frac{i\pi}{N}})\psi_{n-1}(i\pi
\!-\!2\theta\!+\!{\textstyle\frac{(2-n)i\pi}{N}})({\textstyle\frac{N}{2}
}\!-\!{\textstyle\frac{N\theta}{i\pi}}\!-\!n\!+\!2),
\ee
which is exactly the recurrence we had for the $SU(N)/SO(N)$ case.

In order to calculate fully the minimal $K$-matrices for the two cases,
$SU(N)/SO(N)$ and $SU(N)/S\!p(N)$, we need to use the above recurrence
relation to calculate an expression for $\rho_n(\theta)$ in terms of
basic quantities. Starting from the recurrence
\be
\tilde{\rho}_n(\theta)=\rho(\theta\!+\!{\textstyle\frac{(n-1)i\pi}{N}})
\tilde{\rho}_{n-1}(\theta\!-\!{\textstyle\frac{i\pi}{N}})\psi_{n-1}(i\pi
\!-\!2\theta\!+\!{\textstyle\frac{(2-n)i\pi}{N}})({\textstyle\frac{N}{2}
}\!-\!{\textstyle\frac{N\theta}{i\pi}}\!-\!n\!+\!2)
\ee
we can prove by induction that the following expression for
$\rho_n(\theta)$ holds:
\be
\rho_n(\theta)=\Omega(n)\prod_{k=1}^{n}{\rho(\theta\!+\!{\textstyle
\frac{(2k-n-1)i\pi}{N}})}\prod_{k=2-n}^{n-2}{(\eta(i\pi\!-\!2\theta\!+\!
{\textstyle\frac{2ki\pi}{N}}))^{\lfloor\frac{n-|k|}{2}\rfloor}}
\ee
where
\be
\eta(\theta)={\textstyle\frac{N\theta}{2i\pi}}\omega{(\theta)}.
\ee

We note that the procedure we have used to calculate the minimal
$K$-matrices is valid only for $n\leq\frac{N}{2}$. We can obtain the
boundary scattering description for the particles of rank $>\frac{N}{2}$
by recalling that they are the conjugates of particles of rank
$<\frac{N}{2}$ and fusing together conjugate vector particles. The
minimal $K$-matrix for the conjugate vector particle is given
by~\cite{macka01}
\be
K_{\bar{1}}(\theta)=\rho(\theta)E^{\dag}.
\ee
Thus, more generally, we can exchange the $K$-matrix of a particle
and that of its conjugate particle by hermitian conjugation of the
$P_n$ or $P_n^{(n-2r)}$ operators.

\subsection{The PCM $K$-matrices}

Now that we have calculated the minimal $K$-matrices for the
cases $SU(N)/SO(N)$ and $SU(N)/S\!p(N)$ we need to determine the CDD
factors (which will have the same functional form, in terms of the basic
vector particle CDD quantities, for both cases). We have the following
recurrence for the CDD factors
\be
Y_n(\theta)=\res{\phi=\frac{ni\pi}{N}}{X_{(1,n-1)}(\phi)}Y_1(\theta\!+\!
{\textstyle\frac{(n-1)i\pi}{N}})\tilde{X}_{(n-1,\bar{1})}(2\theta\!+\!
{\textstyle\frac{(n-2)i\pi}{N}})\tilde{Y}_{n-1}(\theta\!-\!{\textstyle
\frac{i\pi}{N}}),
\ee
where
\be
\tilde{Y}_{n-1}(\theta)=\frac{Y_{n-1}(\theta)}{\res{\phi=\frac{(n-1)i\pi
}{N}}{X_{(1,n-2)}(\phi)}}\,,\quad\tilde{X}_{(n-1,\bar{1})}(\theta)=
\frac{X_{(n-1,\bar{1})}(\theta)}{\res{\phi=\frac{(n-1)i\pi}{N}}{X_{(1,n-
2)}(\phi)}}.
\ee
From this we can prove inductively the result
\be
Y_n(\theta)=\res{\phi=\frac{ni\pi}{N}}{X_{(1,n-1)}(\phi)}\prod_{k=1}^{
n}{Y_1(\theta\!+\!{\textstyle\frac{(2k-n-1)i\pi}{N}})}\prod_{k=2-n}^{n-2
}{(X_{(1,\bar{1})}(2\theta\!+\!{\textstyle\frac{2ki\pi}{N}}))^{\lfloor
\frac{n-|k|}{2}\rfloor}}.
\ee
Having calculated the complete PCM $K$-matrices we can move on to
analyse their pole structures.

\section{The physical strip pole structure of the PCM $K$-matrices and
boundary Coleman-Thun mechanisms}
\label{sec:pkb}

We recall the structure of the PCM $K$-matrices
\be
K_n^{PCM}(\theta)=Y_n(\theta)\Big({K_n(\theta)}_L\otimes{K_n(\theta)}_R\Big).
\ee
Determining the physical strip poles of these $K$-matrices involves
collecting together all the pole and zero locations of the constituent
parts, remembering that any poles and zeroes in the minimal
$K$-matrices, $K_n(\theta)$, will appear with their order doubled as
there are two copies present in the above expression. We present just
the results here as the method is only a matter of meticulous
accounting.

In section~\ref{subsec:hsv} we gave four possible CDD factors for the
vector particle $K$-matrix
\be
Y_1(\theta)=(\gamma N\!+\!2)(\delta N\!+\!4),
\ee
where $\gamma,\delta=1,3$ were the four possibilities. The vector
particle scattering indicates no preference between these four choices.
The higher rank particle scattering, however, singles out a preferred
choice in each of the cases $SU(N)/SO(N)$ and $SU(N)/S\!p(N)$.

For $SU(N)/SO(N)$ we find that $\gamma\!=\!3,\delta\!=\!1$ is the
preferred choice. The other three possibilities lead to physical strip
poles which are either inconsistent (for example a pole at
$\theta=\frac{i\pi}{2}$ when $K_n(\frac{i\pi}{2})$ does not project onto
a scalar representation subspace of $S\!p(N)$) or undesirable. With the
preferred choice $K_n^{PCM}(\theta)$ has no physical strip poles, and
only has the following simple zeroes on the physical strip
\be
\theta={\textstyle\frac{i\pi}{2}},\ \theta={\textstyle\frac{i\pi}{2}}
\!+\!{\textstyle\frac{(1-n)i\pi}{N}},\ \theta={\textstyle\frac{i\pi}{2}}
\!-\!{\textstyle\frac{ni\pi}{N}}\quad({\rm not\ present\ when}\ n={
\textstyle\frac{N}{2}}).
\ee
Since there are no poles on the physical strip in $K_n^{PCM}(\theta)$
for any $n\leq\frac{N}{2}$, we conclude that no boundary bound states can
be formed. We have a consistent set of $K$-matrices describing boundary
interactions in the case of boundary conditions corresponding to the
symmetric space $SU(N)/SO(N)$ with no boundary bound states.

For $SU(N)/S\!p(N)$ we find that $\gamma\!=\!3,\delta\!=\!1$ is again
allowed and produces the, rather trivial, structure described above.
More interestingly, $\gamma\!=\!\delta\!=\!1$ is also allowed (again the
other two possibilities lead to inconsistent/undesirable physical strip
poles) and this we take as the preferred choice. With this choice
$K_n^{PCM}(\theta)$ has the following poles and zeroes which lie on the
physical strip
\bea
\theta={\textstyle\frac{i\pi}{2}}\quad(n\ {\rm odd\ only}),\ \theta=
{\textstyle\frac{i\pi}{2}}\!+\!{\textstyle\frac{(1-n)i\pi}{N}}\qquad{\rm
simple\ zeroes}, \nonumber \\
\theta={\textstyle\frac{i\pi}{2}}\quad(n\ {\rm even\ only})\qquad{\rm
simple\ pole}, \\
\theta={\textstyle\frac{i\pi}{2}}\!+\!{\textstyle\frac{(2-n)i\pi}{N}},
\ \theta={\textstyle\frac{i\pi}{2}}\!+\!{\textstyle\frac{(4-n)i\pi}{N}},
\ldots,\ \theta=\left\{\begin{array}{cc}
{\textstyle\frac{i\pi}{2}}\!-\!{\textstyle\frac{2i\pi}{N}} & n\ {\rm
even} \\
{\textstyle\frac{i\pi}{2}}\!-\!{\textstyle\frac{i\pi}{N}} & n\ {\rm odd}
\end{array}\right.{\rm double\ poles}. \nonumber
\eea
We note that at rapidity $\theta=\frac{i\pi}{2}\!+\!\frac{(2r-n)i\pi}{N}
$ for $r=1,2,\ldots,\lfloor\frac{n}{2}\rfloor$ the minimal $K$-matrix
projects onto the subspace associated with
\be
P_n^{(n-2r)}\oplus P_n^{(n-2r-2)}\oplus\dots\oplus P_n^{(b)}\,,\qquad
{\rm where}\ b=\left\{\begin{array}{cc}
0 & n\ {\rm even} \\
1 & n\ {\rm odd}
\end{array}\right..
\ee

\subsection{Boundary Coleman-Thun mechanisms for $SU(N)/S\!p(N)$}

Having established the physical strip pole structure we can examine it
to see what bulk-boundary couplings are required for a consistent
description of the boundary interactions.

Firstly we note that $K_n^{PCM}(\theta)$ has a simple pole at
$\theta=\frac{i\pi}{2}$ when $n$ is even. Since the boundary is in the
ground state, appealing to lemma 2 of a paper by Mattsson and
Dorey~\cite{matts00}, there can be no explanation for such simple poles
other than a coupling of the even rank bulk particles to the boundary at
this rapidity. The minimal $K$-matrix contains only a contribution from
the projector $P_n^{(0)}$ at $\theta=\frac{i\pi}{2}$ and so projects
onto the embedded $S\!p(N)$ scalar representation subspace. This is
consistent with such a coupling of the bulk particles to the boundary
and we conclude that the even rank particles couple to the boundary at
$\theta=\frac{i\pi}{2}$. Diagrammatically we have \\
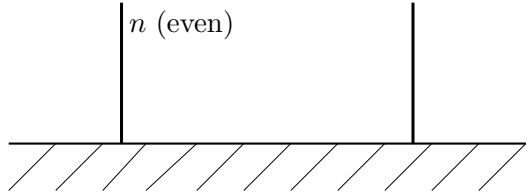
\begin{pspicture}(-4.65,0)(6.875,3)
\psline(0,.625)(6.875,.625)
\psl(0,0)(.625,.625)
\psl(.625,0)(1.25,.625)
\psl(1.25,0)(1.825,.625)
\psl(1.825,0)(2.5,.625)
\psl(2.5,0)(3.125,.625)
\psl(3.125,0)(3.75,.625)
\psl(3.75,0)(4.375,.625)
\psl(4.375,0)(5,.625)
\psl(5,0)(5.625,.625)
\psl(5.625,0)(6.25,.625)
\psl(6.25,0)(6.875,.625)
\pslv{.04}(1.5,.625)(1.5,2.5)
\pslv{.04}(5.375,.625)(5.375,2.5)
\rput(2.3,2.2){{\small $n$ (even)}}
\end{pspicture} \\
\indent We are now in a position to explain all the other physical strip poles,
namely the double poles at $\theta=\frac{i\pi}{2}\!+\!\frac{(2r-n)i\pi}{
N}$ for $r=1,2,\ldots,\lfloor\frac{n-1}{2}\rfloor$, by boundary
Coleman-Thun mechanisms. We consider the diagram \\
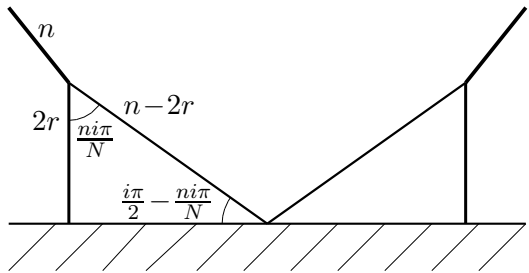
\begin{pspicture}(-4.65,0)(6.875,4)
\psline(0,.625)(6.875,.625)
\psl(0,0)(.625,.625)
\psl(.625,0)(1.25,.625)
\psl(1.25,0)(1.825,.625)
\psl(1.825,0)(2.5,.625)
\psl(2.5,0)(3.125,.625)
\psl(3.125,0)(3.75,.625)
\psl(3.75,0)(4.375,.625)
\psl(4.375,0)(5,.625)
\psl(5,0)(5.625,.625)
\psl(5.625,0)(6.25,.625)
\psl(6.25,0)(6.875,.625)
\pslv{.05}(0,3.5)(.8,2.5)
\pslv{.04}(.8,.625)(.8,2.5)
\pslv{.03}(.8,2.5)(3.4375,.625)
\pslv{.03}(6.075,2.5)(3.4375,.625)
\pslv{.04}(6.075,.625)(6.075,2.5)
\pslv{.05}(6.075,2.5)(6.875,3.5)
\psl(2.8375,.625)(2.8459,.725)
\psl(2.8459,.725)(2.8718,.825)
\psl(2.8728,.825)(2.9179,.925)
\psl(2.9179,.925)(2.9485,.9726)
\rput(2.1,.925){{\small $\frac{i\pi}{2}\!-\!\frac{ni\pi}{N}$}}
\psl(.8,2)(.9,2.0101)
\psl(.9,2.0101)(1,2.0427)
\psl(1,2.0417)(1.1,2.1)
\psl(1.1,2.1)(1.2,2.2)
\rput(1.15,1.75){{\small $\frac{ni\pi}{N}$}}
\rput(.5,3.2){{\small $n$}}
\rput(.5,2){{\small $2r$}}
\rput(2,2.2){{\small $n\!-\!2r$}}
\end{pspicture} \\
which provides a valid bCTm for the double pole in $K_n^{PCM}(\theta)$
at $\theta=\frac{i\pi}{2}\!+\!\frac{(2r-n)i\pi}{N}$ as the diagram is
second order. We recall that the order of a diagram is given by
\be
{\rm order}=\#{\rm internal\ edges}-2\#{\rm closed\ loops}
\ee
(there is no contribution from $K_{n-2r}^{PCM}(\frac{i\pi}{2}\!-\!\frac{
ni\pi}{N})$ to the order as this is finite, non-zero). \\
The diagram we have drawn is valid for all $2\leq n<\frac{N}{2}$. In the
case $n=\frac{N}{2}$ it becomes \\
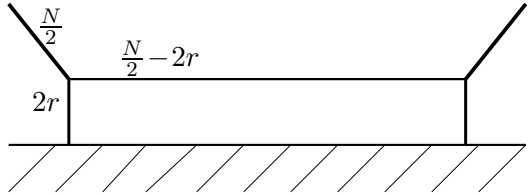
\begin{pspicture}(-4.65,0)(6.875,3)
\psline(0,.625)(6.875,.625)
\psl(0,0)(.625,.625)
\psl(.625,0)(1.25,.625)
\psl(1.25,0)(1.825,.625)
\psl(1.825,0)(2.5,.625)
\psl(2.5,0)(3.125,.625)
\psl(3.125,0)(3.75,.625)
\psl(3.75,0)(4.375,.625)
\psl(4.375,0)(5,.625)
\psl(5,0)(5.625,.625)
\psl(5.625,0)(6.25,.625)
\psl(6.25,0)(6.875,.625)
\pslv{.05}(0,2.5)(.8,1.5)
\pslv{.04}(.8,.625)(.8,1.5)
\pslv{.03}(.8,1.5)(6.075,1.5)
\pslv{.04}(6.075,.625)(6.075,1.5)
\pslv{.05}(6.075,1.5)(6.875,2.5)
\rput(.56,2.2){{\small $\frac{N}{2}$}}
\rput(.5,1.2){{\small $2r$}}
\rput(2,1.78){{\small $\frac{N}{2}\!-\!2r$}}
\end{pspicture} \\
which is still second order. Since we are dealing with representation
conjugating $K$-matrices a factor $K_{\frac{N}{2}-2r}^{PCM}(0)$ must be
present in the above or the diagram would not be consistent. We allow
this since we can make the $\frac{N}{2}\!-\!2r$ bulk particle
arbitrarily close to the boundary.

Thus we have constructed a consistent picture of boundary interactions
in which all physical strip poles are explained by the coupling of all
even rank bulk states to the boundary at rapidity
$\theta=\frac{i\pi}{2}$.

\section{Conclusions}

We have investigated the scattering of particles off the boundary in the
$SU(N)$ Principal Chiral Model on a half-line with conjugating boundary
conditions. We have constructed boundary scattering matrices for all the
bulk particles in the two cases where the BCs correspond to the
symmetric spaces $SU(N)/SO(N)$ and $SU(N)/S\!p(N)$. Having examined the
physical strip pole structures of these $K$-matrices, we have concluded
that none of the bulk particles couple to the boundary in the case where
the BCs correspond to $SU(N)/SO(N)$. Taking BCs corresponding to
$SU(N)/S\!p(N)$ we have found that all even rank bulk particles couple
to the boundary at rapidity $\theta=\frac{i\pi}{2}$, whilst all other
physical strip poles in the $K$-matrices can be explained by boundary
Coleman-Thun mechanisms involving just these bulk-boundary couplings.

Thus we have revealed that there are no boundary bound states in either
case of conjugating boundary conditions. This is in contrast to the
model with non-conjugating boundary conditions, which appears to have a
very rich boundary spectrum and is the subject of current study by the
author.

\vspace{0.2in} {\bf Acknowledgements}
\vspace{1ex}

\noindent I would like to thank Niall Mackay for discussions and
corrections to the draft manuscript. I thank also Gustav Delius for
helpful discussions and the UK EPSRC for a Ph.D. studentship.


\parskip 8pt
\baselineskip 15pt {\small
\begin{thebibliography}{99}

\bibitem{lusch78}
M. Luscher, {\em Quantum non-local charges and the absence of particle
production in the 2D non-linear $\sigma$-model}, Nucl. Phys. {\bf
B135} (1978) 1

\bibitem{berna91}
D. Bernard, {\em Hidden Yangians in 2D massive current algebras},
Commun. Math. Phys. {\bf 137} (1991) 191

\bibitem{evans99}
J. M. Evans, M. Hassan, N. J. MacKay and A. Mountain, {\em Local
conserved charges in principal chiral models}, Nucl. Phys. {\bf
B561} (1999) 385, {\tt hep-th/9902008}

\bibitem{ogiev87}
E. Ogievetsky, N. Reshetikhin, P. Wiegmann, {\em The principal chiral
field in two dimensions on classical lie algebras: the Bethe-ansatz
solution and factorized theory of scattering}, Nucl. Phys. {\bf B280} (1987)
45-96

\bibitem{dorey98}
Patrick Dorey, {\em Exact $S$-matrices}, {\tt hep-th/9810026}

\bibitem{macka01}
N. J. MacKay, B. J. Short, {\em Boundary scattering, symmetric spaces
and the principal chiral model on a half-line}, {\tt hep-th/0104212}

\bibitem{deliu01}
G. W. Delius, N. J. MacKay, B. J. Short, {\em Boundary remnant of
Yangian symmetry and the structure of rational reflection matrices},
Phys. Lett. {\bf B522} (2001) 335, {\tt hep-th/0109115}

\bibitem{matts00}
Peter Mattsson and Patrick Dorey, {\em Boundary spectrum in the sine-Gordon
model with Dirichlet boundary conditions}, J. Phys. {\bf A33} (2000)
9065-9094, {\tt hep-th/0008071}

\bibitem{ghosh93}
S. Ghoshal and A. Zamolodchikov, {\em Boundary S-matrix and
boundary state in two-dimensional integrable quantum field
theory}, Int. J. Mod. Phys. {\bf A9} (1994), 3841, {\tt
hep-th/9306002}

\bibitem{dorey98b} Patrick Dorey, Roberto Tateo and Gerard Watts, {\em
Generalisations of the Coleman-Thun mechanism and boundary reflection
factors}, {\tt hep-th/9810098}

\end{thebibliography}}
\end{document}